\documentclass{pasj00}
\draft

\begin{document}
\SetRunningHead{Author(s) in page-head}{Running Head}

\title{Grisms Developed for FOCAS}

\author{
Noboru \textsc{Ebizuka,} \altaffilmark{1, 2}
 Koji S. \textsc{Kawabata,} \altaffilmark{3}
 Keiko \textsc{Oka,} \altaffilmark{4, 2}
 Akiko \textsc{Yamada,} \altaffilmark{4}
 Masako \textsc{Kashiwagi,} \altaffilmark{4}
 Kashiko \textsc{Kodate,} \altaffilmark{4}
 Takashi \textsc{Hattori,} \altaffilmark{5}
 Nobunari  \textsc{Kashikawa,} \altaffilmark{6}
and Masanori \textsc{Iye} \altaffilmark{6}
}

\altaffiltext{1} {Plasma Nanotechnology Research Center, Nagoya University, 1 Furo-cho, Chikusa-ku, Nagoya, Aichi 464-8603} 
\altaffiltext{2} {RIKEN (The Institute of Physical and Chemical Research), 2-1 Hirosawa, Wako, Saitama  351-0198}\email{ebizuka@riken.jp}
\altaffiltext{3} {Hiroshima Astrophysical Science Center, Hiroshima University, 1-3-1 Kagamiyama, Higashi-Hiroshima, Hiroshima 739-8526}
\altaffiltext{4} {Department of Mathematical and Physical Science, Japan Women's University, 2-8-1 Mejirodai, Bunkyo-ku, Tokyo 112-8681}
\altaffiltext{5} {Subaru Telescope, National Astronomical Observatory of Japan, 650 North A'ohoku Place, Hilo, HI 96720,  USA}
\altaffiltext{6} {National Astronomical Observatory of Japan, 2-1-1 Osawa, Mitaka, Tokyo 181-8588}


%

\KeyWords{faint object --- instrumentation: high dispersion grism --- techniques: high diffraction efficiency} 

\maketitle

\begin{abstract}
Faint Object Camera and Spectrograph (FOCAS) is a versatile common-use optical instrument for the 8.2m Subaru Telescope, offering imaging and spectroscopic observations.  FOCAS employs grisms with resolving powers ranging from 280 to 8200 as dispersive optical elements.  A grism is a direct-vision grating composed of a transmission grating and prism(s).  FOCAS has five grisms with replica surface-relief gratings including an echelle-type grism, and eight grisms with volume-phase holographic (VPH) gratings.  The size of these grisms is 110 mm$\times$106 mm in aperture with a maximum thickness of  110 mm.  We employ not only the dichromated gelatin, but also the hologram resin as a recording material for VPH gratings.  We discuss the performance of these FOCAS grisms measured in the laboratory, and verify it by test observations, and show examples of astronomical spectroscopic observations.
\end{abstract}

\section{Introduction}
\begin{figure} [ht]
  \begin{center}
    \includegraphics [width=54mm, clip]{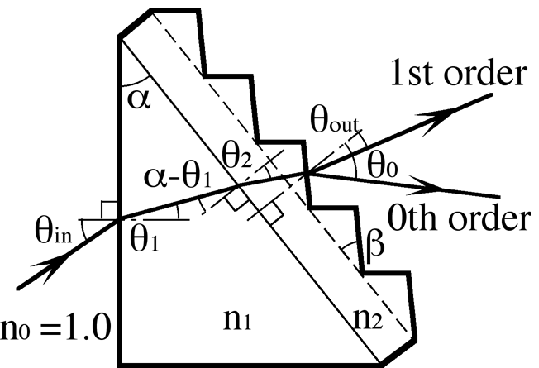}
    \includegraphics [width=60mm, clip]{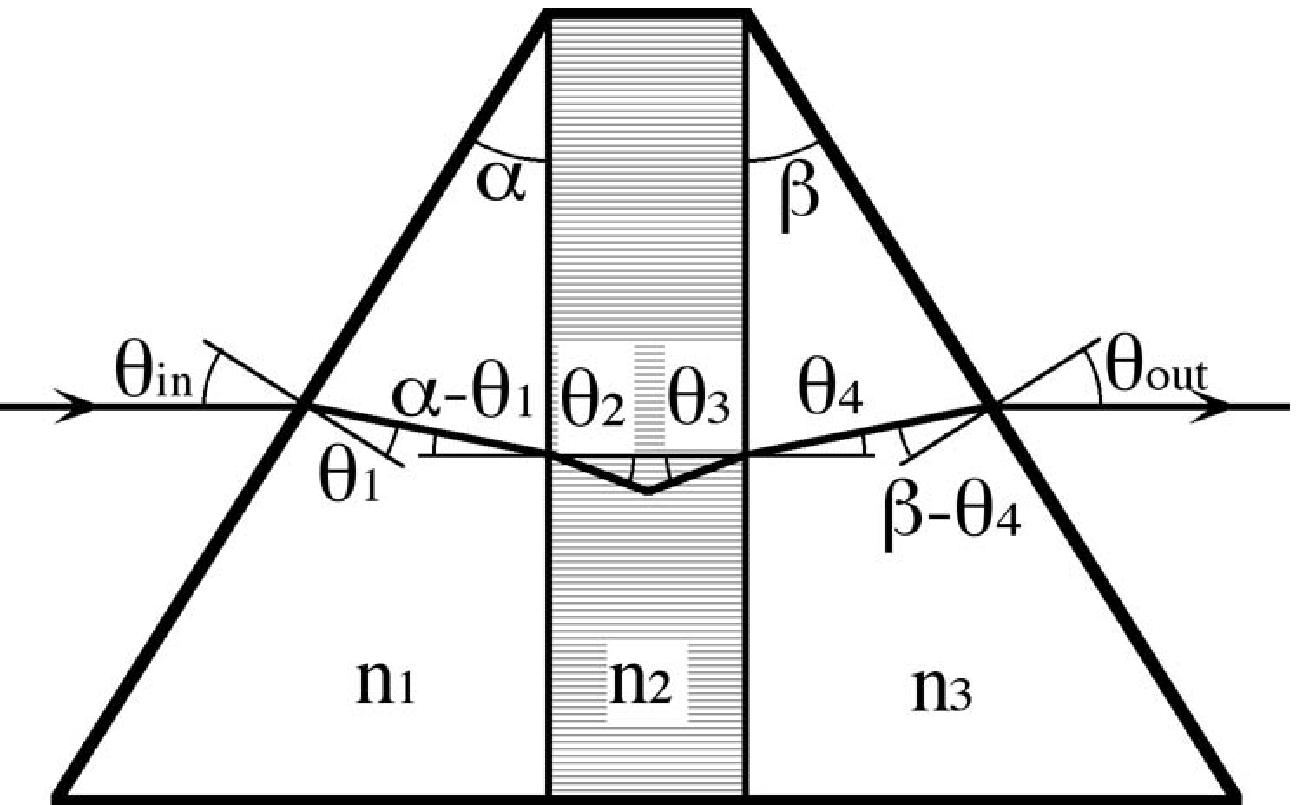}
   \end{center}
  \caption{Light propagation of a surface relief grism (left) and a VPH grism (right).}\label{fig1}
\end{figure}

As shown in figure \ref{fig1}, a grism is a direct-vision dispersive element composing of a transmission grating and prism(s).  A grism is designed to cancel the diffraction angle of a transmission grating by the refracting angle of the prism(s); the dispersed beam of the central wavelength of the specific diffraction order goes straight forward in parallel with the incident beam.  This has a great advantage of an astronomical instrument in that we can switch promptly between the imaging mode and the spectroscopic mode, just by setting a slit onto the focal plane of a telescope and inserting a grism into the collimated beam section of the instrument.

Faint Object Camera and Spectrograph (FOCAS) is an optical instrument for the Cassegrain focus of the 8.2m Subaru Telescope on Mauna Kea, Hawaii (Kashikawa, et al. 2002).  FOCAS has been operated as an open common-use instrument since 2001 April.  A pair of 4K$\times$2K back-side illuminated CCD imagers of FOCAS with pixels 15 $\mu$m square cover the entire Cassegrain field of view, 6' in diameter, at a pixel scale of 0".103/pixel.  FOCAS offers several observing modes: direct imaging, long-slit spectroscopy, multi-slit spectroscopy, and slitless spectroscopy.  Each of these modes can be used together with polarimetric elements.  As dispersive elements, FOCAS provides five surface relief grisms, including an echelle type grism, and eight volume-phase holographic (VPH) grisms optimized for various resolving powers ($R=\lambda/\Delta\lambda$), ranging from 280 to 8200 with various central wavelengths designed so as to cover the entire spectral region.  A direct-vision prism is also installed as a cross disperser for the echelle-type grism.

We describe the specifications and performance of these grisms in section 2 and 3 that are now in use for FOCAS.  Examples for astronomical spectroscopic observations are also presented in section 4.  We present conclusions in section 5.

\section{Development of Grisms}
Because of the limitation of the collimated-beam size of FOCAS and the size specification for housing grisms in the grism turret, the physical size of a FOCAS grism is specified to be 110 mm$\times$106 mm in aperture with a maximum thickness 110 mm along the optical axis, so as to offer an effective aperture of 102 mm in diameter.

\begin{table}
  \caption{Specifications of FOCAS surface-relief grisms.}\label{tab1}
  \begin{center}
    \begin{tabular}{lclclclc|c|c|c|}
      \hline
      Name & Modulation & Nominal & Wavelength & Resolving & Prism & Vertex Angle \\
      & frequency & central wave- & coverage  & power\footnotemark[$*$] & material & of prism \\
      & [Periods/mm] & length [nm] & [nm] &  ($R=\lambda/\Delta\lambda$) &($n_d$) & [$^{\circ}$] \\ \hline
      Very Low & 75 & 650 & 433$\sim$867 & 280 & S-FSL5 (1.49) & 5.75\\ \hline
      Low & 150 & 650 & 433$\sim$867 & 570 & S-FSL5  (1.49)& 11.55\\
      2nd && 340 & 340$\sim$510 &&&\\ \hline
      Middle Blue & 300 & 550 & 367$\sim$733 & 1030 & S-FSL5 (1.49)& 19.7\\ \hline
      Middle Red & 300 & 750 & 500$\sim$1000 & 1430 & S-BSL7 (1.52)& 26.1\\
      2nd && 390 & 360$\sim$540 & 1570 &&\\ \hline
      Echelle 2nd & 175 & 972 & 780$\sim$1050 & 2740 & S-FSL5 (1.49)& 45.0\\ 
      3rd & & 655 & 560$\sim$780 & 2770 & &\\
      4th & & 498 & 440$\sim$560 & 2810 & &\\
      5th & & 404 & 370$\sim$440 & 2840 & &\\ \hline
      Cross & & 435 & & & PBM2Y (1.62)& 19.0\\
      Disperser & & & & & S-FSL5Y (1.49)& 25.0\\
      \hline
       \multicolumn{4}{@{}l@{}}{\hbox to 0pt{\parbox{85mm}{\footnotesize    
     \par\noindent
     \footnotemark[$*$] Calculated for a 0".4 slit.
    }\hss}}
    \end{tabular}
  \end{center}
\end{table}

\subsection{Surface-Relief Grisms for the First Diffraction Order}
As shown in the left panel of figure \ref{fig1}, grooves of surface-relief grisms for FOCAS are of sawtooth shape, called an echelette grating.  As the vertex angle, $\alpha$, and the refractive index, $n_1$, of a prism increases, the angular dispersion of a grism becomes large.  However, the angular dispersion of a surface-relief grating is limited by the ratio of the groove spacing period, $\Lambda$, to a specified wavelength, $\lambda$.  The diffraction properties of a surface-relief grating can be evaluated mainly by using geometrical and wave optics.  The diffraction efficiency of a surface-relief grating of the first diffraction order diminishes rapidly within $\Lambda/\lambda\lesssim4$.  Employing the electromagnetic theory, such as rigorous coupled wave analysis (RCWA), is necessary for properly evaluating the diffraction properties of a surface-relief grating with $\Lambda/\lambda\lesssim10$ (Oka et al. 2003).  We have fabricated surface-relief grisms of 720 and 600 grooves/mm with optical glass prisms designed for central wavelengths at 510 and 670 nm, respectively.  The central wavelengths are coincident with the blazed wavelengths caluclated by geometrical optics, and the efficiencies of the grisms evaluated by wave optics were expected to be $\gtrsim 70\%$ at the central wavelengths.  However, the measured efficiencies of the grisms were 48\% and 45\% at the central wavelengths because the groove period of the grisms were $\Lambda/\lambda=2.7$ and 2.5, respectively.  These efficiencies are consistent with numerical calculation through RCWA.  Furthermore, the angular dispersion of a replicated grism with a high index prism is limited by the critical angle at the boundary between the prism and the low-index resin replica.  For these reasons, the surface-relief grisms for the first diffraction order having $\Lambda/\lambda\gtrsim4$ are used for FOCAS.  That is, the surface-relief grisms are used for applications with comparatively lower resolving power, $R \lesssim1700$.  We decided that we should employ VPH gratings for the high-dispersion grisms of the first diffraction order for FOCAS (Ebizuka et al. 2003).

\begin{figure} [ht]
  \begin{center}
    \includegraphics [width=36.45mm, clip]{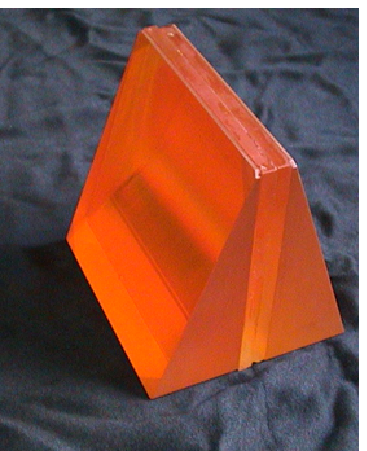}
    \includegraphics [width=113.88mm, clip]{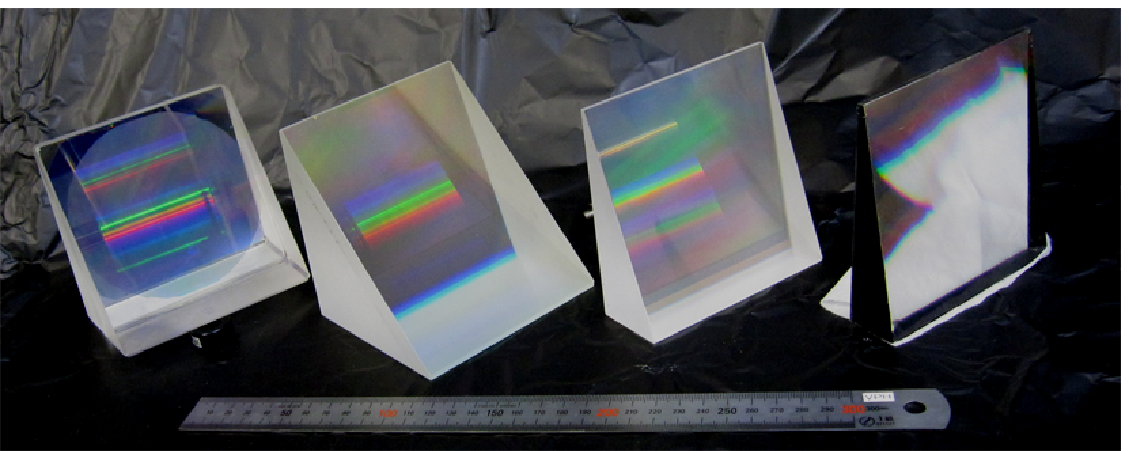}
   \end{center}
  \caption{FOCAS grisms.  From left to right: VPH grisms with ZnSe prisms and with optical glass prisms, and replicated surface-relief grisms of echelle, Middle Red and Low.}\label{fig2}
\end{figure}

\begin{figure} [ht]
  \begin{center}
    \includegraphics [width=80mm, clip]{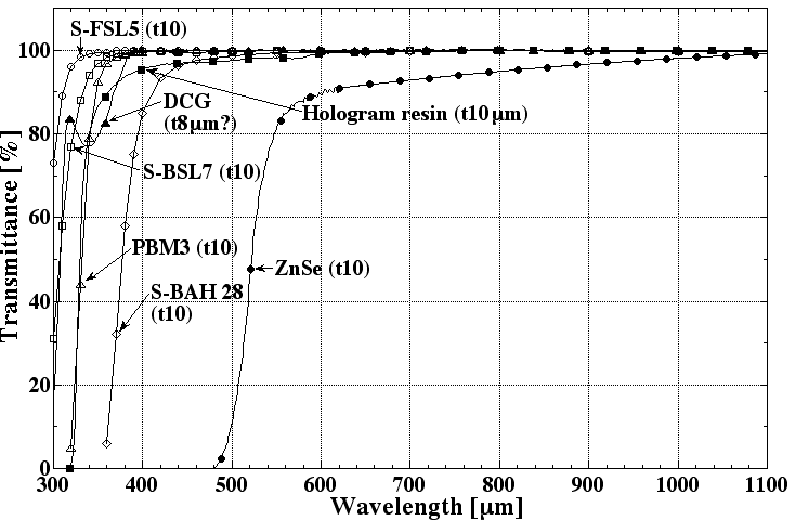}
    \includegraphics [width=80mm, clip]{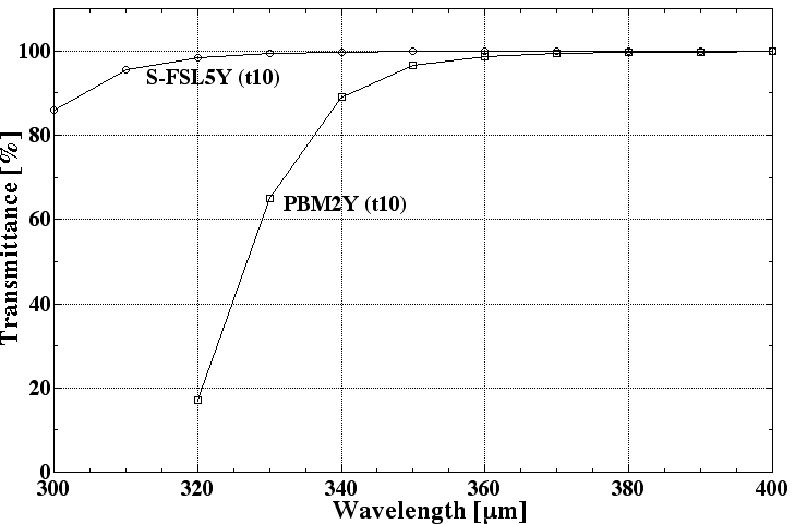}
   \end{center}
  \caption{Internal transmittance curves of optical glasses, ZnSe, hologram resin, and DCG (dichromated gelatin) for grisms in the optical region (left).  Internal transmittance curves of optical glasses for a cross disperser in the UV region (right).}\label{fig3}
\end{figure}

\begin{figure} [ht]
  \begin{center}
    \includegraphics [width=80mm, clip]{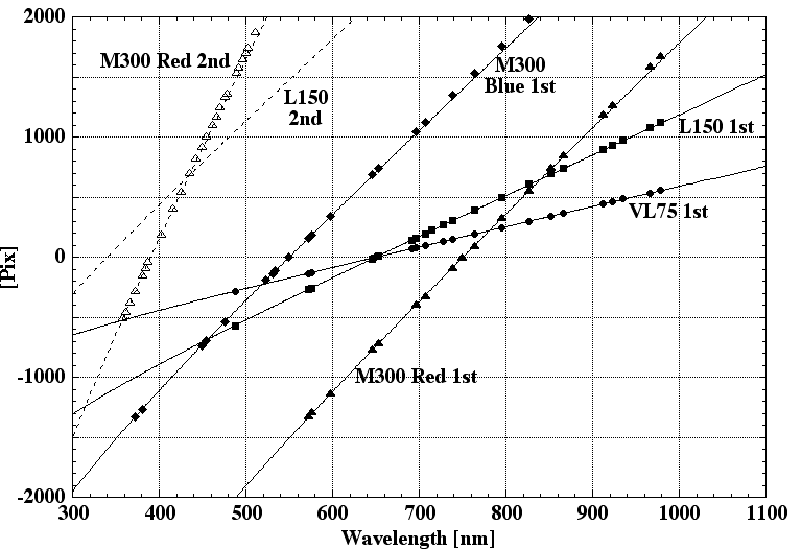}
    \includegraphics [width=80mm, clip]{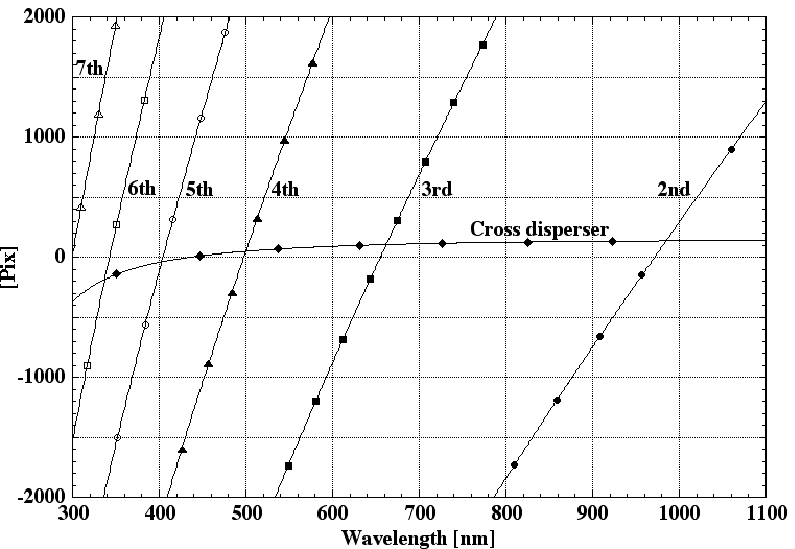}
   \end{center}
  \caption{Spectral positions of the four surface-relief grisms (left), the echelle grism and the cross disperser (right) on the CCD imager of FOCAS.  Solid and dashed lines indicate spectral positions calculated from the designed specifications with the central slit, and symbol marks along the lines show the spectral position measured by using a Th-Ar hollow cathode lamp in the left panel.  Solid lines with symbol marks indicate spectral positions calculated from the designed specifications with the central slit in the right panel.}\label{fig4}
\end{figure}

Four surface-relief grisms for the first diffraction order, manufactured by replication directly onto a prism substrate of an optical glass, are now in use in FOCAS.  Two of them are shown in figure \ref{fig2} far right.  The back surface of the replica of the Very Low grism is tilted $\sim3^{\circ}$ by mounting in an inclined cell; also, the surface of prisms for the other surface-relief grisms are tilted with $5^{\circ}$ so as to avoid ghosts of any surface reflection.  We have not found any apparent ghost patterns in actual spectroscopic data, even for bright objects, such as flux-standard stars.  As shown in table \ref{tab1}, the surface-relief grisms have diffraction gratings with groove density ranging between 75 and 300 grooves /mm, providing $R= 280\sim1430$ for a slit width of 0".4.  The optical glass of the prism substrate for the Middle Red grism is S-BSL7 (n$_d$: 1.52)\footnote{n$_d$: refractive index at the helium d line (587.56 nm).}; the glass for the other grisms is S-FSL5 (n$_d$: 1.49), synthesized by Ohara Inc.  S-BSL7 and S-FSL5 are $\gtrsim 95\%$ transmission for thickness of 10 mm for wavelengths longer than 350 and 320 nm, respectively (figure \ref{fig3}, left).  The grating of the Middle Blue grism was replicated by Jobin Yvon Inc., and gratings of the other grisms were replicated by Richardson Grating Laboratory Inc. (RGL).  Solid lines and dashed lines in the left panel of figure \ref{fig4} show the designed spectral positions of the four surface-relief grisms on the CCD imager of FOCAS.

\subsection{Echelle Grism and Cross Disperser}
\begin{figure} [ht]
  \begin{center}
    \includegraphics [width=80mm, clip]{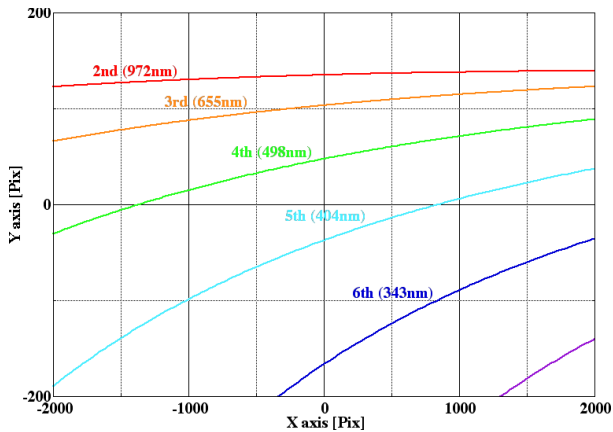}
   \end{center}
  \caption{Echelle spectrum format on the CCD imager.  The direction of the cross dispersion is magnified by $\sim7$ times.  The central wavelengths of each diffraction order is put in parentheses.}\label{fig5}
\end{figure}

An echelle grism, composed of a diffraction grating of higher diffraction orders with a groove density of 175 grooves/mm and a prism substrate of S-FSL5 with a vertex angle of $45^{\circ}$ (figure \ref{fig2}, center), was designed for direct-vision wavelengths of the 2nd diffraction order at 972 nm, the 3rd order at 655 nm, the 4th order at 498 nm and the 5th order at 404 nm with $R= 2740\sim2840$ (table \ref{tab1}). The back surface of the reprica of prism is also tilted by $5^{\circ}$.  Manufacturing the master grating by means of a ruling-engine and making replica of the echelle grism were performed by RGL.  The cross disperser of a direct-vision prism of an octagonal shape with a physical size of 120 mm$\times$120 mm for the echelle grism, which is another direct-vision dispersive element, consists of optical-glass prisms of S-FSL5Y (n$_d$: 1.49, $\nu_d$: 70.3)\footnote{$\nu_d$:Abbe's number (indicator for refractive index dispersion) at the helium d line.} with a vertex angle of $25^{\circ}$ and PBM2Y (n$_d$: 1.62, $\nu_d$: 36.3) with a vertex angle of $19^{\circ}$ (table \ref{tab1}).  S-FSL5Y and PBM2Y have $\gtrsim 95\%$ transmission for a thickness of 10 mm in longer wavelengths than 310 nm and 350 nm, respectively (figure \ref{fig3}, right).  Solid lines with symbol marks in the right panel of figure \ref{fig4} shows the designed spectral positions of the echelle grism and the cross disperser.  Figure \ref{fig5} shows the echelle spectrum format for the combination of the echelle grism and the cross disperser on the FOCAS CCD imager.  It is noted that the second diffraction order of the echelle grism is available for long-slit spectroscopy mode with R=2740 by the combined use of an appropriate filter.

\subsection{VPH Grism}
A VPH grating has a layer, several times the grating spacing period in thickness, with a sinusoidal modulation of the refractive index.  A calculation of the diffraction properties for a VPH grating requires making ample use of the electromagnetic theory, because the property depends on the strength of electromagnetic coupling between the light wave and the modulation.  The modulation amplitude, the thickness of the layer, and the incident angle of the light should be carefully designed so as to achieve strong electromagnetic coupling between the light wave and the grating layer.  The VPH grating achieves a very high diffraction efficiency of up to 100\% for s- or p-polarized light at the specific wavelength, when the incident and diffracted beams satisfy the Bragg condition. (Kogelnik 1969; Barden et al. 1998, 2000; Rallison et al. 2003).  However, the efficiency properties of a VPH grating for s- and p-polarized light (the electric fields parallel and orthogonal to the grating lattice respectively) are different.  The peaks of the efficiencies for s- and p-polarized light slightly shift to shorter and longer wavelength, respectively, from the peak wavelength of the non-(or $45^{\circ}$- or circular-) polarized light (Dickson et al. 1994; Baldry et al. 2004).  VPH gratings with an average refractive index of 1.54 and the index modulation of 0.02 in semiamplitude provide diffraction efficiencies of nearly 100\% for s-polarized light within lattice period between 0.7 and 4 times of wavelength (Oka et al. 2003).  As shown in the right panel of figure \ref{fig1}, a VPH grism basically consists of a VPH grating sandwiched between two prisms, so that the designed central wavelength is aligned parallel to the incident beam.  Since the incident angle from a prism to a VPH grating, $\alpha-\theta_1$, is smaller than the vertex angle, $\alpha$, of the prism when the incident beam of the grism is parallel to the base surface of the prism --- $\theta_{\mathrm{in}}=\alpha$ in the right panel of figure \ref{fig1} --- the limitation of angular dispersion for a VPH grism by the critical angle is less restricted than that of a surface-relief grism is.  Furthermore, the VPH grism splits a prism into two symmetrical pieces.  These properties show that a VPH grism has a capability for applications with a higher resolving power than that of a surface-relief grism in its first-diffraction order (Ebizuka et al. 2003, 2011).

\begin{table}
  \caption{Specifications of FOCAS VPH grisms.}\label{tab2}
  \begin{center}
    \begin{tabular}{lclclclc|c|c|c|}
      \hline
      Name & Modulation & Nominal central & Wavelength & Resolving & Prism & Vertex angle \\
      & frequency & wavelength & coverage  & power\footnotemark[$\dagger$] & material & of prism \\
      & [Periods/mm] & [nm] & [nm] &  ($\lambda/\Delta\lambda$) &(n$_d$)& [deg.] \\ \hline
      VPH 450 & 1000 & 450 & 380$\sim$520 & 3100 & PBM3 (1.61)&20.0 \\ 
      VPH 520 & 990 & 520 & 450$\sim$590 & 3400 & S-BAH28 (1.72) & 20.0 \\
      VPH 650\footnotemark[$*$] & 665 & 650 & 530$\sim$770 & 2770 & PBM3 (1.61)& 20.0 \\
      VPH 680 & 1570 & 690 & 650$\sim$730 & 8200 & ZnSe (2.61)& 20.0 \\
      VPH 800\footnotemark[$*$] & 1320 & 800 & 750$\sim$850 & 7370 & ZnSe (2.61)& 20.0 \\
      VPH 850 & 364 & 790 & 560$\sim$1020 & 1690 & S-BSL7 (1.52)&16.0 \\
      VPH 900 & 560 & 900 & 750$\sim$1040 & 3010 & S-BAH28 (1.72)& 20.0\\
      VPH 950\footnotemark[$*$] & 1110 & 950 & 890$\sim$1010 & 6940 & ZnSe (2.61)& 20.0 \\
      \hline
       \multicolumn{4}{@{}l@{}}{\hbox to 0pt{\parbox{85mm}
       {\footnotesize    
     \par\noindent
     \footnotemark[$*$] Hologram recording material is DCG.
     \par\noindent
     \footnotemark[$\dagger$] Calculated for 0".4 slit.
    }\hss}}
    \end{tabular}
  \end{center}
\end{table}

Table \ref{tab2} summarizes the specifications of eight VPH grisms developed for use in FOCAS, two of which are shown left most and left in figure \ref{fig2}.  Three out of the eight VPH grisms (VPH650, VPH800 and VPH950 grisms) used dichromated gelatin (DCG) as a hologram recording material fabricated by RALCON Inc.  We fabricated VPH gratings for the other five VPH grisms (VPH450, VPH520, VPH680, VPH850 and VPH900 grisms) by using a hologram resin supplied by Nippon Paint Co. Ltd. (Ebizuka et al. 2003; Kashiwagi et al. 2004; Nakajima et al. 2008).

\begin{figure}
  \begin{center}
    \includegraphics [width=120mm, clip]{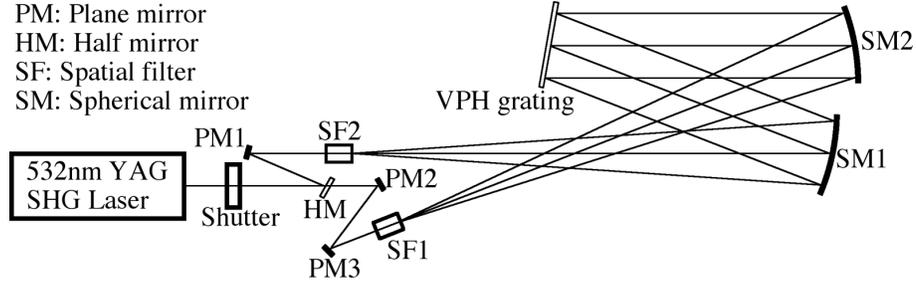}
   \end{center}
  \caption{Schematic representation of an exposure optics of a two-beam interferometer with spherical mirror collimators for a holographic grating.}\label{fig6}
\end{figure}

The hologram resin consists of a radical polymerized monomer (RPM), a cation polymerized monomer (CPM), a dye, a solution and a few other trace elements (Kawabata et al. 1994).  The resin with glass beads is spread on a glass substrate, and the resin is covered with another glass substrate after heating for vaporization of the solution in the resin.  The thickness of the resin's layer is thus controlled by the diameter of the glass beads as a spacer.  Two kinds of optical glasses, 6 mm in thickness, are used for substrates of hologram plates of the VPH gratings, namely S-FSL5 for VPH450 grism and S-BSL7 for the other VPH grisms.  Figure \ref{fig6} shows a schematic representation of a two-beam interferometer used to generate a hologram with a 532 nm green laser.  The optical layout cancels any aberrations of the two spherical mirrors used as collimators.  RPM of the hologram resin polymerized by UV and visible light has a higher refractive index, while CPM polymerized by UV light has a lower refractive index.  When bright and dark stripes of interferometric laser modulation are formed onto a hologram plate by the interferometer, RPMs are polymerized in the bright parts.  As the concentration of RPM is decreased in the bright parts, the RPMs diffuse from the dark to bright parts and the CPMs diffuse from the bright to dark parts.  The modulation of a volume phase hologram is thus formed by the interferometric laser exposure (figure 6).  The modulation is fixed, while both of CPM and residual RPM resins are polymerized by UV light exposure after laser exposure (Ebizuka et al. 2003).  We note that the development of the DCG is a wet process, while that of the hologram resin is a dry one.

\begin{figure} [ht]
  \begin{center}
    \includegraphics [width=80mm, clip]{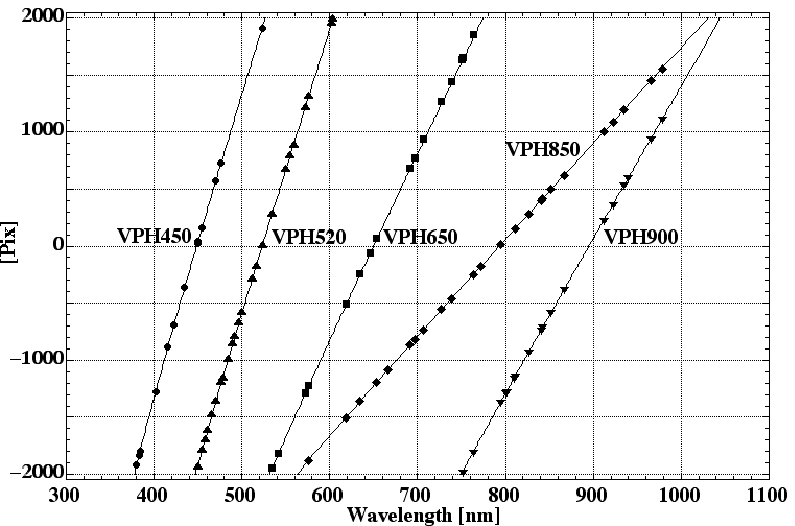}
    \includegraphics [width=80mm, clip]{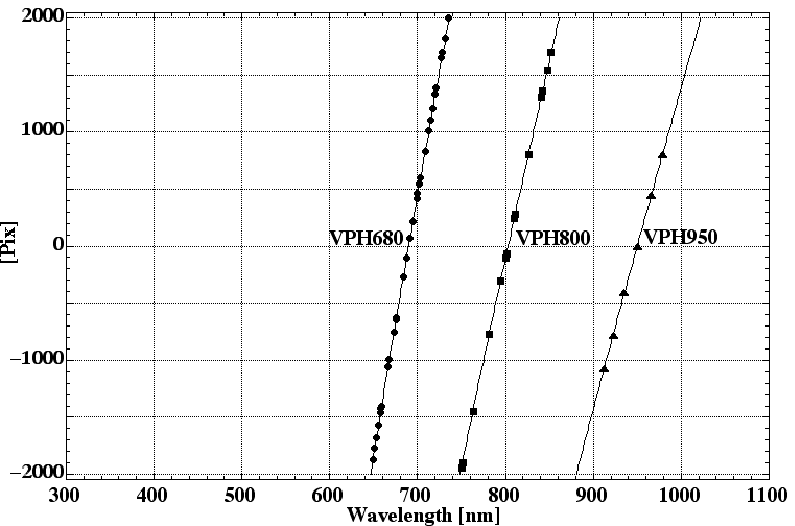}
   \end{center}
  \caption{Dispersion of VPH grisms.}\label{fig7}
\end{figure}

The DCG is able to achieve large modulation amplitude, as high as 20\% of average refractive index, while it is difficult to fabricate a DCG layer above 20$\mu$m in thickness.  On the other hand, the hologram resin achieves the modulation amplitude up to 6\% of average refractive index, while thickness of the hologram resin is easily controlled by diameter of glass beads.  The modulation amplitude of the hologram resin is not proportional to the exposure time, because the development of the hologram is processed under exposure, and the modulation must be affected by diffracted light of itself.  Since the band width of a grism becomes wider as the modulation amplitude of a hologram layer becomes larger, the DCG has an advantage for a VPH grism of large angular dispersion with a wide band width.  Both the DGC and the hologram resin were available for any FOCAS VPH grisms because the specifications of band width were achieved by the modulation below 0.04 in semiamplitude.  The hologram resin is appropriate for VPH850 grism because the grism needs the thickness of the hologram layer to be above 20$\mu$m to achieve a high efficiency of nearly 100\%.

The VPH grating and two prisms were assembled into a VPH grism by an optical adhesive.  Five VPH grisms among eight employed prisms of conventional optical glasses: namely, PBM3 (n$_d$: 1.61), S-BAH28 (n$_d$: 1.72) and S-BSL7; the remaining three VPH grisms used zinc selenide (ZnSe, n$_d$: 2.61) prisms (table \ref{tab2}).  The glass prism had an antireflection coating for the air surface, and the ZnSe prism had antireflection coatings for both the air and the adhesive surfaces.  Note that PBM3 and S-BAH28 are $\gtrsim 95\%$ transmission in longer wavelengths than 360 and 440 nm, respectively, and ZnSe is $\gtrsim 85\%$ transmission in longer wavelengths than 580 nm (figure \ref{fig3}, left).  PBM3 was used for the VPH 650 grism because large glass blanks for prisms with high refractive index except PBM3 were necessary to be delivered nearly a half year in the middle of 2002.  The solid lines in figure \ref{fig7} show the designed spectral positions of the VPH grisms on the CCD imager of FOCAS.

The VPH450 grism had originally been designed for the central wavelength being at 400 nm.  The central wavelength, vertex angle, and substrate of prism were changed from 400 to 435 nm, from $22^{\circ}.5$ to $20^{\circ}$, and from S-FSL5 to PBM3, respectively, while the specification of the VPH grating did not change by a communication error.  Finally the central wavelength of the grism became 450 nm.  VPH850 grism was designed for intermediate dispersion observation ranging  between 600 to 1000 nm.  The nominal central wavelength of the grism is 800 nm, but it was called the VPH850, because another VPH grism with ZnSe prisms had been already called the VPH800 grism.

The peak wavelength could be changed by using an offset slit, because a peak wavelength of a VPH grism efficiency changes with the incident angle, $\theta_{\mathrm{in}}$.  For examples, VPH450 grism with an offset slit is feasible for an UV (360$\sim$400 nm) observation, and  the combination of the VPH 850 or VPH900 or VPH 950 grism and an appropriate offset slit, in which peak wavelength of the grism efficiency changes to 1000 nm, is suitable for observations of high-redshift objects with $z\gtrsim7$. 

\section{Performance Verification of Grisms}
\subsection{Laboratory Measurements}
\begin{figure} [ht]
  \begin{center}
    \includegraphics [width=80mm, clip]{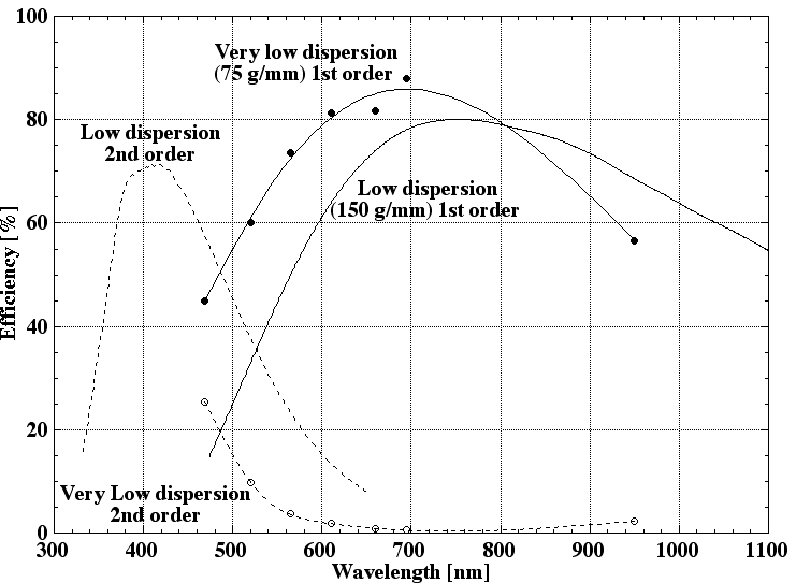}
     \includegraphics [width=80mm, clip]{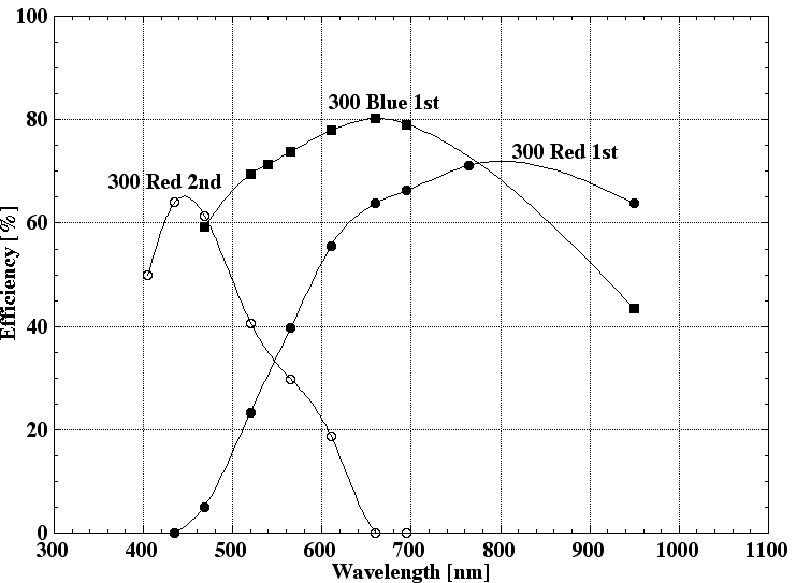}
   \end{center}
  \caption{Efficiency curves of the surface-relief grisms.}\label{fig8}
\end{figure}

\begin{figure} [ht]
  \begin{center}
    \includegraphics [width=80mm, clip]{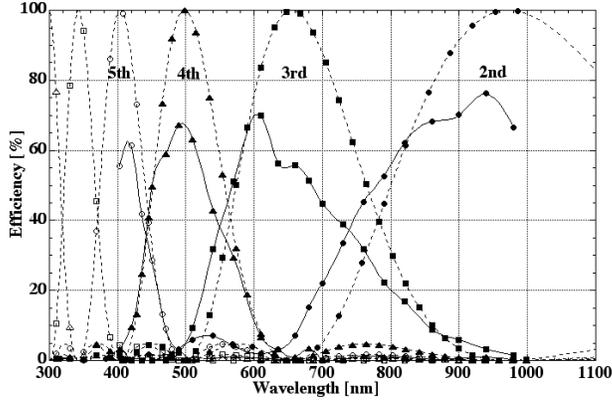}
   \end{center}
  \caption{Efficiency curves of the echelle grism.  Solid lines shows the measured efficiencies, and dashed lines indicate the designed efficiencies calculated by using wave optics}\label{fig9}
\end{figure}

The diffraction efficiencies of grisms were measured by using a spectrophotometer with an integrating sphere, or a grating measurement system. The grating measurement system consists of various LEDs or a monochromator with a stabilized halogen lamp, collimator lens, sample stage, automatic rotational arm, telescopic optics within an exit slit, photovoltaic detector, lock-in amplifier, A/D convertor and PC.  As shown in figure \ref{fig8}, the measured peak efficiencies of the Very Low and Low (the 1st and 2nd diffraction orders) grisms are 85\%, 80\% and 71\% for polarized light at a position angle of $45^{\circ}$, respectively, and these of the Middle Blue and Middle Red (the 1st and 2nd diffraction orders) grisms are 80\%, 71\% and 64\%, respectively, for polarized light at a position angle of $45^{\circ}$.  The wavelength of the peak efficiencies of all grisms shifted toward longer wavelength by 50$\sim$100 nm.  Figure \ref{fig9} shows the measured efficiencies of the echelle grism; the peak efficiencies of the 2nd through the 5th diffraction orders were 76\%$\sim$62\% for polarized light at a position angle of $45^{\circ}$ (Kawabata et al. 2003a).

\begin{figure} [ht]
  \begin{center}
    \includegraphics [width=80mm, clip]{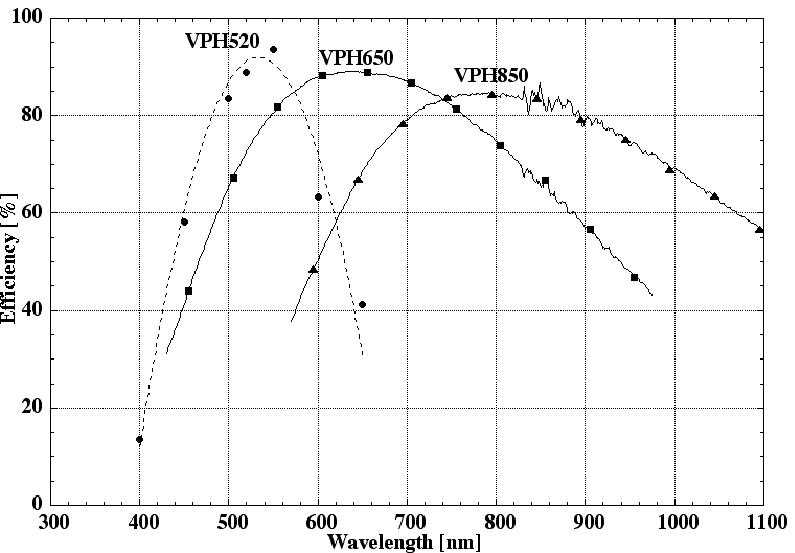}
    \includegraphics [width=80mm, clip]{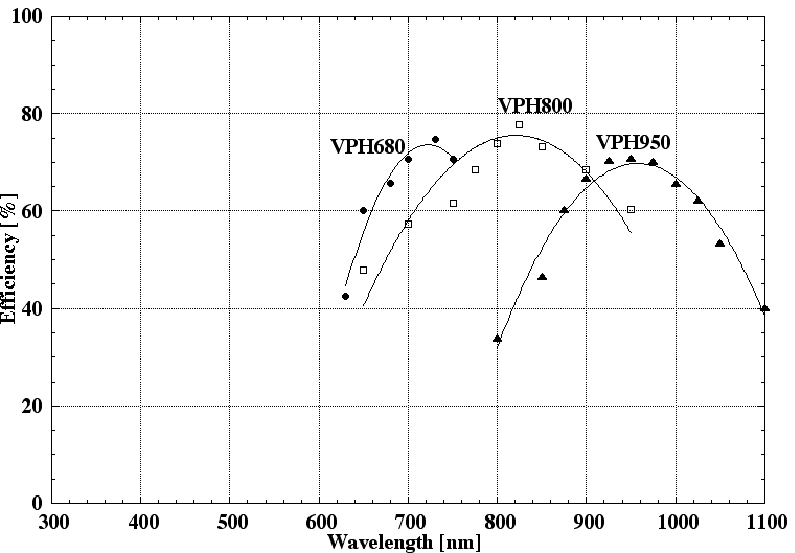}
   \end{center}
  \caption{Efficiency curves of the VPH grisms.}\label{fig10}
\end{figure}

As shown in the left panel of figure \ref{fig10}, the measured peak efficiencies of the VPH520, VPH650, and VPH850 grisms with glass prisms were 92\%, 90\%, and 85\% at the designed central wavelengths for $45^{\circ}$-polarized light, respectively.  The efficiency of the VPH450 grism was 67\% at 440 nm measured by the spectrophotometer without an integrating sphere, while the measured peak efficiency of the grating for VPH450 grism was 87\% at 380 nm.  The efficiency of the VPH900 grism was $\sim$85\% at 900 nm, while the measured peak efficiency of the grism was 93\% at 990 nm, full width at half maximums (FWHM) were $\sim$450 nm, measured before adhesive wad added.  The measurement of the VPH900 grism included some amount of error by means of setting angle precision of the grism.  Moreover, reflection loss of anti-reflection coatings, evaluated up to 1\%, was subtracted by this method because the two prisms with anti-reflection coating on one of the air surface and a glass substrate were used in order to subtract reflections at the glass and air boundaries.

Corresponding curves for the VPH680, VPH800, and VPH950 grisms with ZnSe prisms are shown in the right panel of figure \ref{fig10}; the peak efficiencies were 75\%, 78\%, and 72\% at the peak wavelength for $45^{\circ}$-polarized light, respectively.  Although the measured efficiency curve of the VPH680 grism shifted away from the designed central wavelength, the peak efficiency at 680 nm was estimated (this paragraph).  The peak efficiency of the grating for the VPH680 grism was 91\%; the measured internal transmittance of ZnSe with a thickness of 10 mm was 92.4\% at 680 nm, the value of which was smaller than our expectation of $\sim$95\% evaluated from transmittance of a ZnSe plate with thickness of 1 mm and the table of refractive indices.  The average thickness of ZnSe in the VPH grism was $\sim$37 mm, and thus the peak efficiency of the grism was estimated to be 68\%, while the target of the peak efficiency for the VPH grisms was above 70\%.  Moreover, the efficiencies at the short and long edges of the spectral range were evaluated to be 40\% and 25\%, respectively, while the target efficiencies at both edges were above 60\%.  Though the grating attained the target peak efficiency, unfortunately the VPH680 grism could not reach the target owing to absorption of the ZnSe substrate and the narrow band width.  The FWHM of the efficiency of the VPH680 grism normalized by the central wavelength was $\sim$15\%, while those of the VPH800 and VPH950 grisms were 30$\sim$40\%, even though $\Lambda/\lambda$ of the VPH grisms with ZnSe prisms are comparable.  The semiamplitude of the index modulation of the VPH680 grism was evaluated to be $\sim$0.03 from the band width, while the semiamplitude of the VPH800 and VPH 950 grisms were evaluated to be 0.06$\sim$0.08 (e.g. figure 5 in Baldry et al. 2004).

The tolerance of the wave-front error, $\sigma$, for the FOCAS grism was set to be $\sigma \leq$ 0.5 waves in rms at each designed wavelength so as to ensure minimal optical degradation.  The actual wave-front errors of all these grisms were measured by a GPI interferometer of Zygo Corp., and were shown to be within the specified tolerance.  

\subsection{On-Site Measurement and Test Observations}
The grisms were installed in the grism turret, which was located near the pupil plane of FOCAS, for on-site measurements.  Solid symbol marks in figure \ref{fig7} and the left panel of figure \ref{fig4} represent actual spectral positions provided by each grism on the CCD imager of FOCAS, as verified by using a Th-Ar hollow cathode lamp of the facility for the Subaru Telescope.  The FWHM of spectral line slightly less than 4 pixels, as measured for the Middle Blue and Middle Red, the 2nd diffraction order of the echelle, VPH450, VPH520, VPH 650 and VPH800 grisms using a 0".4 slit, corresponding to 3.88 pixels, indicates that the spectral degradation due to aberration by grism was negligible.  We note that the efficiency of the VPH650 was compensated by the efficiencies of the Middle Blue grism because the reflectivity of the prime mirror of the Subaru Telescope and sky condition were different due to the condition of the test observation for the other VPH grisms.  We conclude that the resolving powers given in table \ref{tab1} for a 0".4 slit are consistent.  The direct-vision wavelengths of these grisms were confirmed to be compatible with their original design by measuring the spectra of the emission lamp, as shown in figure \ref{fig4} and \ref{fig7}, and in table \ref{tab2}.

\begin{figure} [ht]
  \begin{center}
    \includegraphics [width=80mm, clip]{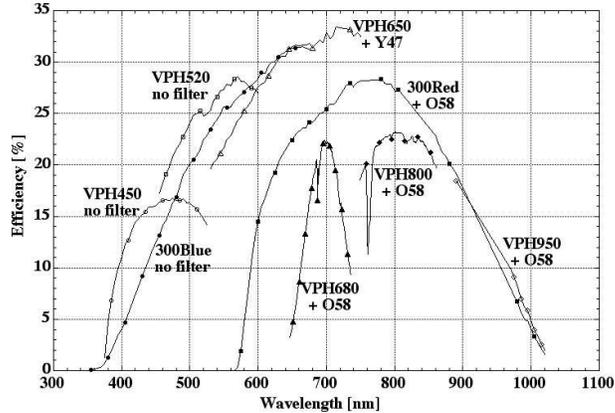}
  \end{center}
 \caption{Total efficiencies of FOCAS grisms including efficiencies of the Subaru Telescope, FOCAS optics, and CCD imager.
 }\label{fig11}
\end{figure}

We obtained the efficiencies of the grisms by observations of spectroscopic standard stars.  Figure \ref{fig11} shows the total efficiency curves of FOCAS grism spectroscopy including throughputs of the Subaru Telescope, the FOCAS optics, and the sensitivities of the CCD imagers, as derived by measuring spectroscopic standard stars and the observed flux corrected for the atmospheric extinction at the zenith.  Considering the throughput of the Subaru Telescope and FOCAS optics, we found that the observed efficiencies are consistent with the laboratory measurements.

\section{Astronomical Observations}
\subsection{Discovery of the Most Distant Galaxies}
FOCAS spectroscopic studies that are aiming to confirm and measure the redshifts of Lyman Alpha Emitter (LAE) candidates, isolated by the SuprimeCam (Miyazaki et al. 2002) narrow-band wide-field surveys, have been highly successful projects.  The echelle grism in the second diffraction order and VPH grisms were mainly used to confirm the characteristic emission line with asymmetric line profiles from galaxies up to 12.9 billion ly away.  LAE surveys at redshifts 5.7 and 6.6 are reported in Shimasaku et al. (2006) and Kashikawa et al. (2006).  Iye et al. (2006) discovered the most distant galaxy at a redshift $z=6.96$, which is a young galaxy, only 800 million years old.  The most recent review is found in Ouchi et al. (2010).  These discoveries led to the finding of redshift evolution in the luminosity function of LAEs, which is interpreted to reflect the last phase of the cosmic reionization (Kashikawa et al. 2006; Ota et al. 2008)

\subsection{Supernovae and GRBs}
Spectroscopy and spectropolarimetry of supernovae (SNe) and gamma-ray bursts (GRBs) can probe the expanding atmosphere and its engine, e.g., chemical abundance, temperatue, density, velocity, and its stucture, including asphericity, as well as the intervening matter as the interstellar and intergalactic media.  The absorpsion/emission lines of SNe are generally broad ($\gtrsim 1000$ km s$^{-1}$).  To achieve both sufficient wavelength resolution and wide wavelength coverage for faint extragalactic SNe, the Middle Red and Middle Blue grisms have been mostly used for these observations.

FOCAS, attached to the Subaru Telescope, has a capability of spectroscopy for distant SNe ($z\gtrsim0.1$) at their early (i.e., photospheric) phases and GRB afterglows at $z\gtrsim5$.  Morokuma et al. (2009) performed follow-up spectroscopy for 39 distant supernova candidates, and confirmed seven probable type-Ia SNe (SNe Ia).  They includes a SNe Ia at $z=1.35$; this is the most distant SN Ia that has been spectroscopically identified by a ground-based telescope to date.  Kawabata et al. (2003b) obtained a spectrum of SN 2003dh, which emerged in associated with the nearby GRB 030329 ($z=0.1685$), at 34-35 d after the GRB.  They suggested that the explosion date of the SN agrees with the GRB in the range $(-8, +2)$ d, based on the similarity in the spectral evolution to the broad-lined SNe Ic ("hypernovae") 1998bw and 1997ef.  Kawai et al. (2006) performed follow-up spectroscopy of GRB 050904; 3 d after the GRB, and confirmed that the spectral cutoff around 900 nm is produced by Ly $\alpha$ absorptions at $z\sim 6.3$ (Totani et al. 2006), which was the most distant GRB discovered before 2008.

\begin{figure} [ht]
  \begin{center}
    \includegraphics [width=120mm, clip]{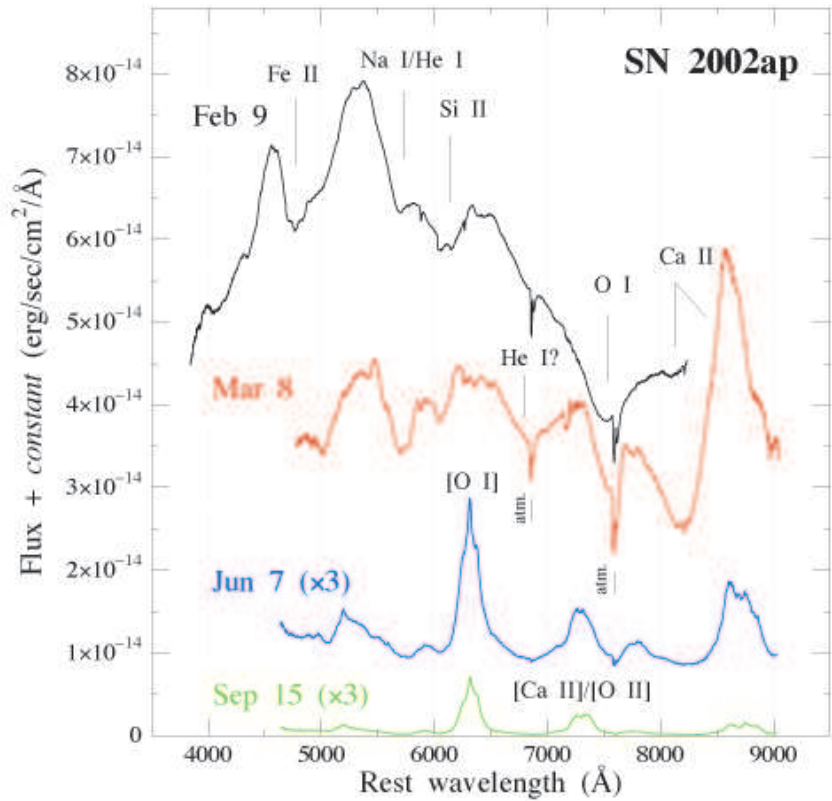}
   \end{center}
  \caption{Spectral evolution of a type-Ic SN 2002ap from 2002 February 9 (around maximum brightness) to September 15 ($\sim$ 220 days after maximum).  The early-phase spectrum is characterized by continuum and broad absorption lines such as O I 7773 and Ca II IR triplet; while the late-phase spectra are dominated by emission lines as [O I] 6300, 6363, [Ca II] 7291, 7323, [O II] 7320, 7330, and the Ca II. These spectra are consistent with mildly energetic explosion with oxygen-rich ejecta of $\sim 2.5$ M$_{\odot}$ and $\sim 4\times 10^{51}$ erg (see Mazzali et al. 2007).
}\label{fig12}
\end{figure}

Subaru/FOCAS also enables us to perform spectroscopic monitoring of nearby SNe through later, nebular phases ($\gtrsim150$ days after the SN explosion, figure \ref{fig12}). Mazzali et al. (2005) obtained a double-peaked [O I]emission line for the non-GRB-associated hypernova SN 2003jd at $\sim 0.9$ yr after the explosion, which suggested that the SN was an aspherical axisymmetric explosion viewed from near the equatorial plane.  This result is interesting because GRBs are believed to be produced from a highly asymmetric explosion; also, some core-collapse SNe are aspherical (and also linked with GRBs).  Maeda et al. (2008) gathered such late-phase spectra of 15 envelope-stripped, core-collapse SNe using FOCAS, and revealed that the high incidence of the double-peaked profiles  are consistent with a model where all envelope-stripped, core-collapse SNe are aspherical. 

About late-phase spectroscopy with FOCAS, a number of case studies have been reported for specific SNe, such as a peculiar SN Ib 2005bf (Tominaga et al. 2005; Maeda et al. 2007), a faint SN Ib 2005cz in an elliptical galaxy (Kawabata et al. 2010), a faint SN Ia 2005hk (Sahu et al. 2008), SN Ib 2008D associated with an X-ray transient (Tanaka et al. 2009).

For the brightest nearby SNe, Subaru/FOCAS enables us to perform high-quality spectropolarimetric observations.  So far, spectropolarimetric results  have been reported, e.g., for SN Ia 2009dc (Tanaka et al. 2010).  They all showed solid intrinsic polarization of continuum light and/or in absorption lines, which can be explained by the global asymmetry of the photosphere, or clumpy distribution of some specific elements within the photosphere, e.g., oxygen, calcium, and silicon.

\subsection{Echo of Emission from Old Supernovae}
Krause et al. (2008a) used FOCAS to obtain the spectrum of the historical supernova SN 1572 observed by Tycho Brahe.  The observation was done in 2008, $\sim$400 yr after direct light of the supernova reached Earth.  This was enabled by observing a "light echo", which is light from the original supernova scattered toward us by a surrounding interstellar cloud.  The Low grism in the second diffraction order and the Middle Blue grism were used together with a 2"-width slit to cover the wavelength range of 380 to 920 nm with a spectral resolution of 2.4 nm (R=160$\sim$380).  The acquired echo spectrum shows the broad absorption and emission features that are commonly observed in supernovae.  A detailed analysis and a comparison of composite spectra with those of the other past supernovae allowed a precise determination of the spectral type, and revealed that SN 1572 belongs to the class of normal type-Ia supernovae.

The same observational technique was also used to determine the type of the Cassiopeia A supernova in 1680. It was found that this supernova was the rare class of type-IIb supernovae, and originated from the collapse of a red supergiant (Krause et al. 2008b).

\subsection{Astronomical Observation with VPH Grisms}
VPH grisms are being use for observations of  the most distant galaxies as mentioned in section 4.1. About other examples, the VPH450 grism was used for observations of a binary system with a compact object (Kubota et al. 2010), and the VPH650 grism was used for observations of a high z star-forming galaxy (Oyama et al. 2004) and of planetary nebulae in a nearby galaxy (Bresolin et al. 2010).  Moreover, any number of observational results of Subaru/FOCAS with VPH grisms will be published in the near future.

\section{Conclusions}
Five surface-relief grisms including the echelle grism were commissioned for common use between 2001 and 2003.  Common use for the VPH450, VPH520, VPH650, VPH680, VPH800 and VPH950 grisms started between 2003 and 2006.  Common use for the VPH850 and VPH900 grisms started in 2010.  The performances have been found to be consistent with the results such as were evaluated in the laboratory, since grisms have been in regular use.  The second diffraction order of the Low and Middle Red grisms are used in short wavelength region in combination with an appropriate filter.  The echelle grism was designed for spectroscopy with wide wavelength coverage; nevertheless, the second diffraction order of the grism is frequently used for a long-slit mode and multislit mode above 800 nm by combination of an appropriate filter.  

Since a VPH grism has a dependence on the slit position and peak-efficiency wavelength, dome flats or twilight flats on observations are important for reduction, especially with multislit mode.  From another viewpoint, the wavelength of the peak efficiency can be adjusted by the slit position.  For examples, the VPH450 grism with an appropriate offset slits is suitable for a UV region (360$\sim$400 nm), and the VPH850 or VPH900 or VPH950 grism with an appropriate offset slit is suitable for observations of galaxies with redshift above 7 (wavelength above 1000 nm).  Although the specifications of the VPH850 grism are similar to that of the Middle Red of surface-relief grism, the efficiency of the VPH grism is higher than that of the surface-relief grism.  This means that a VPH grism is feasible for applications with moderate dispersion.
  
\bigskip
We greatly appreciate M. Kawabata and T. Teranishi of Nippon Paint Co. Ltd. for supplying hologram resin.  The grisms and cross disperser were developed at the request of National Astronomical Observatory of Japan (NAOJ) in collaboration with Communication Research Laboratory (Present name is National Institute of Information and Communications Technology: NICT), RIKEN, Japan Women's University (JWU), Konan University, and Nagoya University.  M. Yoshida of Hiroshima university gave us valuable opinions on grism design.  Numerical calculation for the diffraction properties of the surface-relief grisms were partly performed by T. Sasaki of NAOJ and M. Goto of Max Planck institute.  We thank M. Watanabe, M. Ishikawa, and M. Irisawa and member of Kodate laboratory of JWU for their contributions on VPH grating fabrications and measurements.  Laboratory and on-site measurements were carried out with a great help from FOCAS team, including Y. Ohyama, K. Aoki, Y. Saito, Y. Yadoumaru, H. Taguchi, T. Hashimoto and K. Ota.  N.E. thanks, Y. Suzuki of NICT, A. Makinouchi of RIKEN, F. Kajino of Konan University, S. Sato, and M. Hori of Nagoya University for their supports of the grism studies,and C. Packham of University of Florida for a critical reading of the manuscript. We utilize facility of JWU, RIKEN and the Advanced Technology Center of NAOJ for manufacture of the VPH grisms and for grating measurements.  This research was supported by a grant-in-aid of NAOJ for R\&D of the Subaru Telescope instruments, by a grant-in-aid of RIKEN for practical use of research results, and partially by Grant-in-Aid: for Scientific Research (B), 1998-2000, 09559018; for Specially Promoted Research, 2002-2006, 14002009; and for Scientific Research (S), 2007-2011, 19104004 from the Ministry of Education, Culture, Sports, Science and Technology.

\end{document}